\documentclass[twocolumn,showpacs,preprintnumbers,amsmath,amssymb]{revtex4}

\usepackage{graphicx}

\usepackage{dcolumn}

\usepackage{bm}

\newlength{\figwidth}

\divide\figwidth by 2

\begin{document}




\title{Anomalous spin density distribution on oxygen and Ru in Ca$_{1.5}$Sr$_{0.5}$RuO$_4$:\\ A polarised neutron diffraction study}
\author{ A. Gukasov$^a$, M. Braden $^{a,b,c}$, R.J. Papoular$^a$
S. Nakatsuji$^{d,e}$ and  Y. Maeno$^{d,f}$}
\affiliation{$^a$  Laboratoire L\'eon Brillouin (CEA-CNRS),
CE Saclay, 91191 Gif sur Yvette, France}
\affiliation{$^b$ Forschungszentrum Karlsruhe, IFP,
Postfach 3640, 70621 Karlsruhe, Germany}
\affiliation{$^c$ II. Phys. Inst., Univ. zu K\"oln, Z\"ulpicher Str. 77, 50937 K\"oln, Germany}
\affiliation{$^d$ Departement of Physics, Kyoto University, Kyoto 606-8502, Japan}
\affiliation{$^e$ National High Magnetic Field LAboratory, Tallahassee, Florida 32310, USA}
\affiliation{$^f$ International Innovation Center IIC, Kyoto University, Kyoto
 606-8501, Japan}
\date{\today}
\begin{abstract}

By means of polarized neutron diffraction in a magnetic field of 7.0 T at 1.6 K
an anomalously large magnetization density is observed on the in-plane oxygen
in  Ca$_{1.5}$Sr$_{0.5}$RuO$_4$.  Field-induced  moments of
different ions are determined by refinement on the flipping ratios,
yielding $\mu$$_{Ru}$ = 0.346(11) $\mu$$_B$, $\mu_{O1}$ = 0.076(6) $\mu$$_B$
and $\mu_{O2}$ = 0.009(6) $\mu$$_B$.
The   moment  on the oxygen arises from the strong hybridization
between the Ru-4d and O-2p orbitals.
The maximum entropy magnetization density 
reconstruction reveals a strongly anisotropic density at the Ru site, consistent with the distribution of the
{\it xy}  ($t_{2g}$  band)  {\it d}-orbitals.
\end{abstract}






\maketitle


Layered perovskite ruthenates have attracted considerable interest
because of their magnetic properties and in particular
because of the recent discovery
of superconductivity in  Sr$_2$RuO$_4$  \cite{maeno}.
It is now widely  accepted that superconductivity in this compound is
unconventional
and the pairing  of triplet p-wave symmetry.
Sr$_2$RuO$_4$ does not exhibit any magnetic ordering, but among the related
compounds with a Ru$^{4+}$-ion various magnetic structures are observed :
ferromagnetic order occurs in the metallic perovskite SrRuO$_3$ \cite{srruo3}, antiferromagnetic
one in insulating Ca$_2$RuO$_4$ \cite{ca2ruo4} and
incommensurate order in  Sr$_2$Ru$_{0.91}$Ti$_{0.09}$O$_4$ \cite{ti-condmat}.

According to the band-structure calculations,
the Fermi-surface in Sr$_2$RuO$_4$ is formed by
three bands associated with
the three {\it t$_{2g}$} Ru orbitals, {\it xy,yz} and {\it xz} \cite{mazin1}.
Since {\it 4d} ions generally have more extended {\it d} orbitals than the corresponding ${\it 3d}$
ions, the  {\it 4d}-oxides exhibit larger overlap and hybridization
between the transition metal and O {\it 2p} orbitals.
As a result magnetism becomes more itinerant in character
and there is stronger  interplay
between structural degrees of freedom, and the magnetic and electronic
properties.
Another important point is that in {\it 3d} transition-metal systems,
the strong crystal field interaction dominates over the spin-orbit
interaction and consequently the orbital moment is quenched, $\mu_L=0$, whereas,
in layered ruthenates, much stronger spin-orbit effects may be anticipated,
which should manifest themselves in the radial part of the Ru form factor.

The substitution of Sr by Ca in  Sr$_2$RuO$_4$  leads to 
a complex phase diagram  \cite{ca2ruo4,friedt}.
Already small concentrations of Ca give rise to appearance of a structural distortion characterized
by the rotation of RuO$_6$-octahedra around the c-axis, which is accompanied by a strong  increase of the low temperature
magnetic susceptibility.
For Sr-concentrations higher than 0.5 in Ca$_{2-x}$Sr$_x$RuO$_4$
additional  distortions are
observed  characterized by a tilt of the
octahedra around an in-plane axis. 
In contrast to the rotational distortion, 
the increase of the tilt leads to a decrease of the low temperature magnetic susceptibility.
The critical point x=0.5, where the tilt distortion appears, corresponds to the maximum of 
the low temperature susceptibility. The susceptibility at this point is about 100 times larger than
that of pure Sr$_2$RuO$_4$ \cite{satoru}. We choose this composition for our study,
since the large susceptibility is favorable for the polarized neutron
measurements. 

Polarized neutron diffraction (PND) is a uniquely powerful tool
to study the magnetization densities in crystals, due to the fact that
it provides direct information about the 3D distribution of the magnetization
throughout the unit cell. 
In favorable cases, PND further allows to determine the symmetry of occupied orbitals.
The aim of the present work is two-fold : {\bf i)} to check the presence of unpaired electron
density on the ligands O1 (and O2) which the theory predicts to be anomalously high, and
{\bf ii)} to determine the  radial part of the Ru magnetic form factor and possibly to establish
the symmetry of Ru {\it 4d} orbitals occupied by unpaired electrons.


Polarized neutron measurements were performed on the two-axis lifting-counter
diffractometer {\it 5C1} at the {\it ORPH{\'E}E\,} reactor, LLB CEA/Saclay,
using a wavelength $\lambda$ =0.845\,\AA\ 
(Heusler alloy monochromator, polarization of the incident neutron beam  
$P_0=0.91$).
Higher-order contamination was suppressed to less than 0.01\% by means of
erbium filters.
The flipping ratio (ratio between neutron spin-up and spin-down intensity)
measurements provided the experimental data used in
the subsequent magnetic moment refinement and in the magnetization reconstruction.
The programs MAGLSQ  and SORGAM of the Cambridge Crystallography Subroutine Library
\cite{jane} were used for least squares refinement on the flipping ratios and for
the calculation of magnetic amplitudes, respectively.

A single crystal of  Ca$_{1.5}$Sr$_{0.5}$RuO$_4$  was obtained by a floating
zone technique
\cite{crystal}.
According to Ref. \cite{friedt} Ca$_{1.5}$Sr$_{0.5}$RuO$_4$ crystallizes in the
tetragonal $I4_1/acd$  structure, for which
the oxygen octahedra are rotated around the  $c$ axis with an opposite phase of rotation
for the planes separated by 12.5 \AA . 
The structural parameters of the Ca$_{1.5}$Sr$_{0.5}$RuO$_4$ crystal used in our experiment
can be summarized as follows:
$a$=$b$=5.3\AA, $c$=25.0 \AA , Ru $8a$ position, Ca(Sr) $16d$ z=0.549, O1 $16f$ x=0.193,
 O2 $16d$ z=0.457 (see \cite{friedt} for details).

\begin{figure}[tbp]
\resizebox{0.6\figwidth}{!}{
\includegraphics*{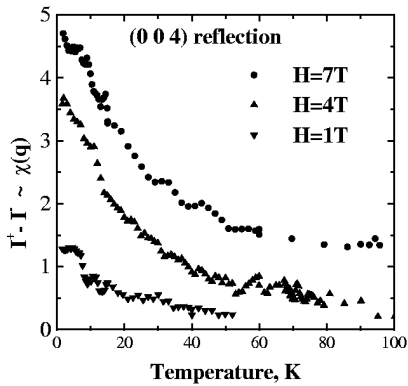}}
\caption{Magnetic susceptibility of Ca$_{1.5}$Sr$_{0.5}$O$_4$
obtained from the polarized neutron intensities of the (004) reflection
with the field parallel to the [110] direction.
\label{fig1}}
\end{figure}

The temperature dependence of the magnetic susceptibility $\chi$
of our Ca$_{1.5}$Sr$_{0.5}$RuO$_4$ crystal was measured directly by means of PND
in magnetic external fields of 1 T, 4 T and 7~T. For this purpose, the spin-up
($I^+$) and spin-down ($I^-$)  intensities
of the (004) reflection were collected as a function of temperature,
and the difference $I^+-I^-$ plotted versus temperature (see  Fig.\ref{fig1}).
This difference,  $I^+-I^-= 2 P_0 \chi H F_N $, where $F_N$ is the nuclear
structure factor for the same reflection and $P_0$ the polarization of the
incoming beam, is proportional to the Fourier component
(004) of the wave-vector dependent susceptibility of the crystal. As seen from
Fig.1 and magnetization data, the susceptibility of
Ca$_{1.5}$Sr$_{0.5}$RuO$_4$ increases 
strongly at low
temperatures but the crystal remains paramagnetic down to 1.6~K;
at  7~T the magnetization is near saturation.

Overall, two sets of polarized neutron flipping ratios were measured at 1.6~K.
Regarding the first, a vertical magnetic field of 7~Tesla was applied
parallel to the  [110]-axis of the crystal and a total of 94
flipping ratios with $\sin \theta /\lambda <0.6$\ \AA$^{-1}$.
This  yielded  53 independent reflections of which only a subset of 13
reflections has a pure oxygen contribution.
(Note that oxygen atoms contribute to all the reflections since they occupy low
symmetry  {\it 16f} and  {\it 16e}  positions.)
The second set consisted  of 97 (34 independent and among them 13 pure oxygen reflections)
flipping ratios also measured at 1.6~K, but with the magnetic field along the
[001] direction.

\begin{figure}[tbp]
\resizebox{0.7\figwidth}{!}{\includegraphics*{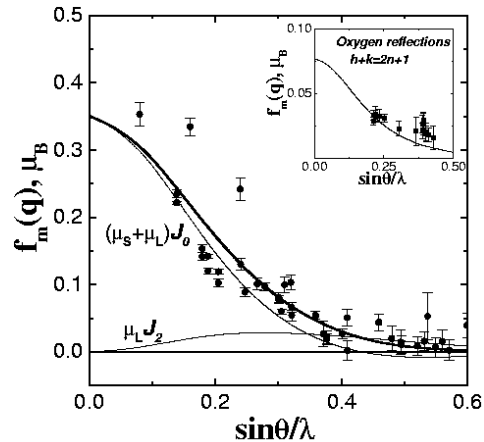}}
\caption{Magnetic amplitudes of Ca$_{1.5}$Sr$_{0.5}$O$_4$ at 1.6 K and 7~T.
Circles represent mixed reflections, normalized by Ru geometrical structure factor.
Squares in the inset are the pure oxygen reflections normalized by the
geometrical structure factor of O1. The solid line is the magnetic amplitude
calculated for the Ru form factor in the dipole approximation.
Thin lines show the decomposition of the latter into its components proportional to $\mu_L$ and $(\mu_S+\mu_L)$ respectively.}
\label{fig2}
\end{figure}

The magnetic amplitudes of all measured reflections were obtained using
the {\it SORGAM} \cite{jane} program, except for those containing a pure oxygen contribution.
They are shown in Fig. \ref{fig2} together with the theoretical magnetic form factor of Ru.
(For convenience, the amplitudes were normalized by the geometrical
structure factor of the Ru.)
As seen from Fig. 2, the majority of measured magnetic amplitudes lie reasonably close
to the theoretical curve.
There are, however, a large number of reflections, especially those at low {\bf Q},
for which the disagreement between experimental and theoretical values is very significant.
A simple analysis shows that these are the reflections with a strong contribution of
the O1 sub-lattice.
Those amplitudes of 16 reflections containing a pure oxygen contribution (h+k=2n+1)
were normalized by the geometrical structure factor of O1.
The  amplitudes calculated in such a way should be proportional to the oxygen form factor,
if one neglects  the aspherical part of Ru magnetic density.
The comparison of experimental results (see inset in Fig.\ref{fig2}) with the solid line
representing the theoretical form factor of oxygen shows that
an additional moment of the order of 0.075 $\mu$$_B$ is present at the O1
site.

Further analysis of the measured flipping ratios was performed by refinement.
The refinement of the first data set using the  tetragonal  group
$I4_1/acd$ demonstrated clearly that there is a very large moment
induced  both on Ru and on O1. The moments induced by the field of 7~T were
obtained as $\mu$$_{Ru}$ = 0.350(15) $\mu$$_B$, $\mu_{O1}$
= 0.076(8) $\mu$$_B$ and $\mu_{O2}$ = 0.014(8) $\mu$$_B$, with $\chi^2=1.87$
for 94 reflections.
For comparison a model in which the oxygen moment is neglected gives
$\mu_{Ru}$ = 0.33(15) $\mu$$_B$ and  $\chi^2=10.2$. The refinement performed
using only 13 pure oxygen reflections yields $\mu_{O1}$ = 0.072(12).
The refinement of the second data set with the applied magnetic field parallel to the [001]
direction gives a very similar result $\mu$$_{Ru}$ = 0.342(15) $\mu$$_B$, $\mu_{O1}$
= 0.076(8) $\mu$$_B$ and $\mu_{O2}$ = 0.009(6) $\mu$$_B$ with $\chi^2=1.98$  for 97 reflections.

These refinements show unambiguously that apart from the relatively large magnetic
density induced at the Ru position a huge amount of density exists at the O1 site.
This moment amounts to more than 20 \% of that of  Ru. Taking the multiplicity of the O1 site
into account, one can conclude that nearly one third of the total magnetic density in the cell
is transferred from Ru to O1. By contrast, the refinement does not show any evidence for the presence
of any magnetic moment neither at the apical oxygen O2 nor at the Sr sites.

\begin{figure*}[t!]
\resizebox{0.75\textwidth}{!}{\includegraphics*{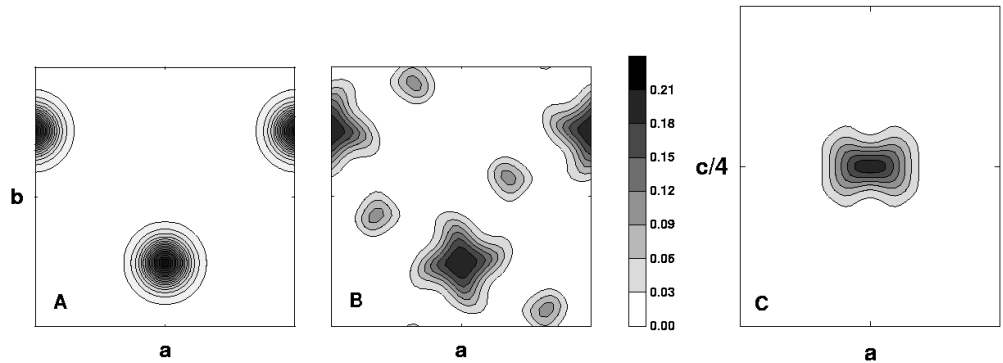} }
\caption{
Maximum entropy reconstruction of the magnetization distribution in
Ca$_{1.5}$Sr$_{0.5}$RuO$_4$ at 1.6 K and 7~T. 
The maps A and B show the sections
passing through the RuO$_2$-plane at 
z=${1\over 8}$ and map C shows  the section perpendicular to the
plane through the Ru-position. 
The edge lengths of the maps correspond to the tetragonal lattice
constant of $a$=5.34\AA ~for maps A and B and to $a$ and a quarter of $c$
for map C respectively.
Note that in space-group I4$_1$/acd a Ru is situated 
at ($1\over 2$,$1\over 4$,$1\over 8$)
and an O1-site lies at (0.19,0.44,$1\over 8$).
The section in A) relates to the adverse 3D non-uniform prior
density used to compute B) and C), which biases the MEM reconstruction against
any magnetization at the O sites as well as against an aspherical magnetization
at the Ru site. The contour step is $0.03 \mu_B\,\AA^{-3}$.
\label{fig3b}
        }
\end{figure*}

In the least-squares analysis, the Ru magnetic form factor in the dipole-approximation was used

${\mu_{Ru}f_{Ru}(Q)=(\mu_S+\mu_L)\langle j_0(Q)\rangle +\mu_L \langle j_2(Q)
  \rangle .}$
Here, $\mu_S$ and $\mu_L$ are the spin and the orbital components of the total  magnetic moment
$\mu_{Ru}$ and $ \langle j_0(Q) \rangle $ and $  \langle j_2(Q) \rangle $ are radial functions.
Upon increasing {\it Q}, $ \langle j_0(Q) \rangle $ decreases whereas 
$  \langle  j_2(Q) \rangle $ has a maximum at {\it Q}=0.3\,\AA$^{-1}$.
We used in the refinement the radial functions of the Ru$^{1+}$ state, since those of higher ionic states
of Ru are not available. At first sight, the utilization of the radial
integral of the ionic state for ruthenates,
being itinerant in nature and often mixed strongly with conduction electrons, is not well justified.
We remind the reader, however, that a similar problem exists in actinide {\it 5f} compounds also
characterized by a strong hybridization. For actinides, it has been found that the radial part
of the wave functions does not change appreciably through the actinide series and for different ionic states.
On the contrary the mixing coefficient $C2=\mu_L/(\mu_S+\mu_L)$  depends strongly on the
hybridization \cite {lander}. The validity of this approach for {\it 4d} compounds remains to be verified
but the description of the Ru form factor by mixing  of $ \langle j_0(Q) \rangle $ and
  $ \langle j_2(Q) \rangle $ of
the Ru$^{1+}$ state can be considered as a good approximation.
The  value of coefficient $C2$ will strongly depend on the {\it ionic} state and
hybridization of Ru, and as in actinides, it can be obtained from refinement against the flipping ratios.
In our instance, the orbital component of the Ru moment was included and was found
to be $\mu$$_L$ = 0.140(15) $\mu$$_B$.
Neglect of the orbital part of the Ru form factor deteriorated the quality of the fit considerably,
from $\chi^2=1.98$ to $\chi^2=2.88$, thereby indicating that the orbital part of the moment exists
and  should be taken into account. The decomposition of the Ru form factor
  into the radial functions 
$ \langle  j_0(Q) \rangle $
and $ \langle j_2(Q) \rangle $ is shown in  Fig.\ref{fig2}.

The results of our refinements show a significant spin density at the O1 site and
it is useful to check to what extent this can be substantiated by a model-free analysis
of our data, obtained for instance by reconstructing the 3D magnetization distribution using the maximum
entropy method (MEM). This method has been shown to give much more reliable results than
conventional Fourier syntheses, by considerably reducing both noise and truncation effects
\cite{papou}.
In order to carry out the MEM reconstructions, the magnetic structure factors of both data sets,
related to two different orientations of the sample, were merged into 70 independent
flipping ratios.
The magnetization density distribution was discretized
into 48$\times$48$\times$240 sections along a,b and c respectively. Then it  was   reconstructed using a conventional {\it uniform} (flat)
density prior \cite{papou}.
Such a procedure is biased against the creation of any magnetic density in the unit cell.
In spite of the negative bias of the MEM procedure the ensuing reconstruction revealed the presence of strong  peaks at the positions of Ru and weaker but
 well pronounced ones at the O1 positions. Numerical integrations of the magnetization density
over the Ru and the O1 positions yield $\mu$$_{Ru}$ = 0.307 $\mu_B$, $\mu_{O1}$ = 0.030$\mu_B$.
Because of the negative bias mentioned above, the MEM values are slightly smaller than those from the
refinement. Moreover, sections through the Ru site revealed a strongly
anisotropic magnetization density,
while sections through the Sr and the O2 sites did not detect any magnetic density above
the $3\times 10^{-3} \mu_B$ level.
To check the robustnest of the density reconstruction
against possible spurious features, the MEM procedure was repeated using  a {\it non-uniform} (adverse)
density prior \cite{devries,papou3}, in which all the magnetisation in the crystal
is spherically concentrated around the Ru sites.
This {\it non-uniform} prior was calculated
{\sl analytically} using  the spherical magnetic density of Ru ions at the $8a$ position via the moment
and radial integrals values obtained from the least-squares refinement mentioned previously.
Such a procedure considerably enhances the  bias against both the creation of magnetic density at
either oxygen sites and the asphericity  of Ru density, as compared to the  {\it uniform} prior strategy.
Nevertheless,  both features of interest, the oxygen density and the asphericity of the Ru distribution,
survive the acid test of an unfavorable {\it non-uniform} prior.
The  final results  using
the {\it non-uniform} prior on the basis of all  measured flipping ratios are
shown in Fig.\ref{fig3b}.





Based on the band-structure calculations a strong hybridization between
Ru-4d and the O-2p orbitals was postulated for the ruthenates in
general \cite{mazin1}. In consequence a significant
oxygen contribution was introduced into the Stoner interaction parameter
for Sr$_2$RuO$_4$.
Since the oxygen may not be polarized in an antiferromagnetic fluctuation,
the Stoner interaction is then q-dependent. Taking this Stoner-interaction into
account the magnetic excitation spectrum for Sr$_2$RuO$_4$ has been calculated
and used as an input for many theories to explain superconductivity in this compound
\cite{mazin1,eremin,japan1,japan2}.
Our finding of a large spin-density
on the oxygen-site validates the basis of these procedures.
So far, there are no calculations of the spin-density induced by an external field
available, but the magnetization in two rather distinct ferromagnetically ordered
ruthenates was calculated \cite{mazin1}, SrRuO$_3$ and Sr$_2$RuYO$_6$.
In both compounds one third of the total spin desity was found on the oxygen
sites, in perfect agreement with the value we observe for the paramagnetic
compound. 
A comparable amount of
transferred magnetization has so far not been observed in any oxide, which
underlines this outstanding property of the ruthenates compared to the
better studied 3d-compounds. For example,  PND found an induced moment of only
a few percent in an Fe-garnet \cite{oxides1} as well as in two manganates \cite{oxides2}.

The anisotropic density
distribution at the Ru is extended along the diagonals of the original
planar perovskite lattice (i.e. in the direction 45 degrees to the bonds)
and it is flattened along the c-direction, see Fig. 3.
This anisotropy suggests that the main
part of the spin-density arrises from the two-dimensional $\gamma$-band
related to the planar {\it xy}-orbitals.
In general the $\gamma$-band is considered as being 
the driving element for a ferromagnetic instability in the 214-ruthenates,
since there is a  van Hove singularity slightly above the Fermi-level \cite{mazin1}.
Fang and Terakura have analyzed the phase diagram of Ca$_{2-x}$Sr$_{x}$RuO$_4$
by ab-initio calculations based on the local density approximation
\cite{fang}. They observe that the rotation of the octahedra flattens and
slightly lowers the $\gamma$-band which enhances the influence of the
van Hove singularity as well as the instability towards ferromagnetism.
The anisotropy observed for Ca$_{1.5}$Sr$_{0.5}$RuO$_4$, a sample
very close to ferromagnetic order, is fully consistent with this
explanation. However, it does not agree with the result of the LDA plus U
calculation by Anisimov et al. for that particular concentration
\cite{anisimov}.

In conclusion, we have unambiguously evidenced  a significant induced spin
density at the positions of
Ru  $\mu_{Ru}$ = 0.350(15) and  anomalously high spin density at the in-plane
oxygen O1  $\mu_{O1}$ = 0.076(8)
$\mu_B$ in Ca$_{1.5}$Sr$_{0.5}$RuO$_4$.
In contrast,
neither the least-squares refinement nor the model-free MEM
procedure show a significant magnetic moment at the sites of Sr and
apical oxygen O2.
The large
magnitude of the moment transferred to O1 is consistent with theorical predictions \cite{mazin1}.

This work was supported by CREST, Japan Science and Technology Corporation,
and by Deutsche Forschungsgemeinschaft through SFB 608.


%

\end{document}